\documentclass[12pt,a4paper,final]{iopart}

\usepackage{iopams}  
\usepackage[breaklinks=true,colorlinks=true,linkcolor=blue,urlcolor=blue,citecolor=blue]{hyperref}

\usepackage[utf8]{inputenc}
\usepackage[T1]{fontenc}
\usepackage{mathptmx}

\usepackage[english]{babel}
\usepackage{hyperref}
\usepackage[pdftex]{graphicx}
\usepackage{amsthm}

\usepackage{tikz}
\begin{document}

\title{Isochrone spacetimes}

\author{Alberto Saa}%
\address{Departmento de Matemática Aplicada, 
 Universidade Estadual de Campinas, \\
 13083-859 Campinas, SP, Brazil}
  \ead{asaa@ime.unicamp.br}

\author{Roberto Venegeroles}
\address{Centro de Matemática, Computação e Cognição,
Universidade Federal do ABC, \\
09210-170 Santo André, SP, Brazil}
\ead{roberto.venegeroles@ufabc.edu.br}

\begin{abstract}
We introduce the relativistic version of the well-known Henon’s isochrone spherical models:   static spherically symmetrical spacetimes in which all bounded trajectories are isochrone in Henon’s sense, { \em i.e.}, their radial periods do not depend on their angular momenta. Analogously to the Newtonian case, these   “isochrone spacetimes”  have as   particular cases the so-called Bertrand spacetimes, in which all bounded trajectories are periodic. We propose a procedure to generate isochrone spacetimes 
by means of an algebraic equation,   present explicitly several families  of these spacetimes,  and discuss
briefly their main properties. We identify, in particular, the family   whose Newtonian limit corresponds to the  Henon’s isochrone potentials  and  that could be considered as the relativistic extension of the original Henon's proposal for the study of globular clusters. Nevertheless, isochrone spacetimes generically violate the weak energy condition and may exhibit  naked singularities, challenging their physical  interpretation in the context of General Relativity. 
\end{abstract}



\section{Introduction}

 The so-called isochrone spherical models were introduced by Michel H\'enon  in the fifties  \cite{MH1,MH2,MH3}  in the study of the dynamics of globular clusters, see  \cite{JBY} for a brief review on the subject.  {Globular clusters
are dense, roughly  spherically symmetric, distributions of stars whose  
  dynamics   is usually described  through an averaged gravitational potential, leading naturally  to the study of general central  potentials. Henon’s isochrone   models are intimately related to
the classical Bertrand's theorem.}
 Strictly speaking, an isochrone
 model in Henon's sense is a Newtonian spherically symmetric gravitational field for which the radial periods of  bounded orbits do not depend on their angular momenta. 
 {The Newtonian   and the harmonic  oscillator potentials}, which according to 
 Bertrand's theorem   are the only central potentials for which all bounded trajectories are periodic (closed),  are also isochrone in
 Henon's sense.   {In fact, 
 Bertrand’s theorem can be considered as  a refinement of the concept of isochrone potential, 
 the Newtonian   and the harmonic  oscillator potentials are the only isochrone potential with
 closed orbits,
 see Section 4.4 of \cite{PPD}.
 }

 In principle,  non-relativistic gravitational configurations
as the    isochrone  spherical  models  
  might be effectively  attained in globular clusters
 as a result of a dynamical mechanism called resonant relaxation, as suggested by H\'enon in his original works \cite{MH1,MH2,MH3},  {see also \cite{Relax,BTGD} for more modern approaches to
 this subject.}
 For further recent developments  on the dynamics of isochrone potentials, see \cite{PPD,PPP,PPC,RP}.  
 The family of  Henon's spherical isochrone potentials includes, besides the  two cases of Bertrand's theorem,
 three other central potentials, namely 
 the so-called H\'enon potential
\begin{equation}
\label{he}
V_{\rm He}(r) = -\frac{k}{b+\sqrt{b^2 + r^2}},  
\end{equation}
and, respectively, the bounded and hollowed potentials 
\begin{eqnarray}
\label{bo}
V_{\rm bo}(r) &=& \frac{k}{b + \sqrt{b^2 - r^2}},  \\
\label{ho}
V_{\rm ho}(r) &=& -\frac{k}{r^2}\sqrt{ r^2- b^2 },
\end{eqnarray}
where $b$ and $k$ are positive constants. It is clear that  the
potentials (\ref{bo}) and (\ref{ho}) are not defined for all $r$ and that the 
 Newtonian potential arises from the
limit $b\to 0$ of $V_{\rm He}$ or $V_{\rm ho}$. The isochrone potentials (\ref{he}) and (\ref{ho}) are asymptotically 
Newtonian for large $r$,  while (\ref{he}) and (\ref{bo}) are effectively 
harmonic near the center
$r=0$.

Some years ago, Perlick   \cite{Perlick} 
 introduced the notion of a Bertrand spacetime, namely a spherically symmetric and
static  spacetime in which any   bounded  trajectory of test bodies is periodic, clearly
extending  the classical Bertrand's theorem   to the realm of General Relativity. 
From  Bertrand's theorem,  we have also that
the azimuthal  angle for the Newtonian and harmonic potential cases is given by  $\Theta = \frac{\pi}{\beta}$  with, respectively, $\beta = 1$ and $\beta=2$. We remind that the azimuthal  angle
corresponds to the angular variation between the closest and farthest points to the center
of a bounded trajectory.  
 Rather surprisingly, Perlick showed that it is always possible to construct a static spherically symmetrical spacetime in which all  test body bounded  trajectories
are periodic for any rational value of the parameter $\beta$. 
 We know that
any  spherically symmetric and
static  spacetime can be cast in a spherical Schwarzschild  coordinate system where its metric takes the form
\begin{equation}
\label{metric}
g_{ab}dx^adx^b = -f(r)dt^2 + \frac{ dr^2}{h(r)} + r^2d\Omega^2,
\end{equation}
where $d\Omega^2 = d\theta^2 + \sin^2\theta d\phi^2$ denotes the usual metric on  {the unit sphere} $S^2$. 
There are two types of Bertrand spacetimes.
For the first type, we have
\begin{eqnarray}
\label{bertr1f}
 {\frac{1}{f(r)}} &=&  G+ \sqrt{\kappa+ r^{-2}}, \\
 {h(r)} &=& \beta^2\left(1+\kappa r^2\right),
\label{bertr1h}
\end{eqnarray} 
whereas the second case corresponds to
\begin{eqnarray}
\label{bertr2f}
 {\frac{1}{f(r)}} &=&  G\mp \frac{r^2}{1-Dr^2 \pm \sqrt{(1-Dr^2)^2 - Kr^4}}, \\
 {h(r)} &=& \frac{\beta^2\left((1-Dr^2)^2 - Kr^4\right)}{2\left(1-Dr^2 \pm \sqrt{(1-Dr^2)^2 - Kr^4}\right)},
\label{bertr2h}
\end{eqnarray}
with arbitrary constants $G$, $\kappa$, $D$, and $K$. For an enlightening geometrical interpretation of the Bertrand spacetimes and their relation with the standard harmonic and Newtonian potentials on curved 3-dimensional manifolds, see \cite{esp}. For the Bertrand spacetimes of the first type, the Newtonian limit is
obtained by setting  
$\kappa = 0$ (locally flat condition) and $\beta = 1$ (absence of conical singularity at the origin), 
and clearly corresponds to the $r^{-1}$ interaction. On the other hand, for the second type of spacetimes,
the locally flat condition and the absence of conical singularity requires, respectively, $D=K=0$ 
and $\beta = 2$, and we have eventually the harmonic potential case.

In the present paper, we introduce the relativistic version of the 
Henon's isochrone models,  which we denominate ``isochrone  spacetimes''.  {They
correspond to the static spherically symmetric spacetimes
(\ref{metric}) in which all bounded timelike trajectories are  isochrone in Henon's sense, {\em i.e.}, their radial periods do not depend
on their  angular momenta. As in the non-relativistic Newtonian case \cite{RP}, the isochrony condition
for a  static spherically symmetric spacetime 
 is
equivalent to demand that the relativistic radial action be additively separable in the two pertinent
constants of motion, namely the energy and the angular momentum. However, as we will see, in clear contrast with the
 Newtonian case, such condition is not a very stringent   restriction in the relativistic domain.} Besides the rather trivial extension of the Bertrand spacetimes for
the case of real $\beta$, there are many other families of isochrone  spacetimes. 
We propose a procedure to generate such new  spacetimes by means of an algebraic equation and
present explicitly some families of solutions. In particular,
we present a family  whose Newtonian limit corresponds to Henon’s isochrone potentials (\ref{he}),
(\ref{bo}) and (\ref{ho}).
Some of these spacetimes   are asymptotically flat and others have regular centers in an analogous way
of their Newtonian counterparts and, hence, they could be useful as  a relativistic extension of the original Henon's proposal for the study of globular clusters. 
However,  isochrone spacetimes generically violate the weak energy condition    and may exhibit
naked singularities and, consequently, their physical interpretation in General Relativity is rather challenging. 
In the next section, we will introduce the isochrone spacetimes and present a procedure to generate them by exploring an algebraic equation. We will present explicitly three large families of these spacetimes and discuss
briefly some of their main properties.
The third and last section is devoted to some concluding remarks, including the issue of the violation of the weak
energy conditions and some brief comments about the causal structure of isochrone spacetimes.

\section{The isochrone spacetimes}

The pertinent Lagrangian for the motion of test bodies in a
static spherically symmetrical spacetime with metric (\ref{metric})
reads
\begin{equation}
\label{lagr}
\mathcal{L} = -f  \dot{t}^2 + \frac{\dot{r}^2}{h} +
 r^2\dot\phi^2,
\end{equation}
 {with the dot denoting the usual derivation with respect to the proper time
of the timelike geodesic,}
where we have, without loss of generality due the spherical symmetry, restricted the motion to the equatorial plane.  {The   geodesic equation  derived form (\ref{lagr}) admits  3 constants of 
motion, which are}
\begin{equation}
 {f\dot t} =  1, \quad \ell = r^2 \dot\phi, \quad 
E = -\frac{1}{f} + \frac{\dot r^2}{h} + \frac{\ell^2}{r^2},
\end{equation}
with $\ell$ and $E$ interpreted, respectively, as the trajectory angular momentum and
energy. This is a simple one-dimensional motion problem and the orbit radial period can be determined  from the conserved quantities 
by exploring elementary methods. We have
\begin{equation}
\label{RF}
 T(E,\ell) =   2 \int_{r_{\rm min}}^{r_{\rm max}} \frac{dr}{\sqrt{h(r)}\sqrt{E-U_{ {\ell}}(r)}},
\end{equation}
where $r_{\rm min}$ and $r_{\rm max}$ are the usual return points of the effective potential 
 {\begin{equation}
\label{unov}
U_\ell(r)= - \frac{1}{f(r)}+  \frac{\ell^2}{r^{2}},
\end{equation} 
which}   is expected to have a local minimum at $r_0$ corresponding to the circular orbits. We also
assume $U_{ {\ell}}''(r_0)>0$,  { {\em i.e.}, the dynamical stability of the
circular orbit.} Moreover, all functions here are assumed to be sufficiently smooth. 
The other important quantity  for our purposes in this work is the  azimuthal angle, which is given 
by
\begin{equation}
\label{apsidal}
 \Theta(E,\ell) = {\ell}  \int_{r_{\rm min}}^{r_{\rm max}} \frac{dr}{r^2\sqrt{h(r)}\sqrt{E-U_{ {\ell}}(r)}} 
\end{equation} 
 {
and
corresponds to the angular variation between the closest and farthest points to the center
of a bounded trajectory.  
Notice that the apsidal angle, another common quantity used in this kind of analysis, is defined as the angular variation during
one radial period and, hence, is twice the azimuthal angle.
Although the effective potential (\ref{unov}) is indeed the relativistic counterpart 
of the sum of the Newtonian
gravitational potential  and the centrifugal barrier term, the presence of the factor
$h(r)$ in (\ref{RF}) and (\ref{apsidal}) spoils any other useful analogy here. 
From (\ref{metric}), we see that $h(r)=1$ basically reduces the problem to the Newtonian one,
since in this case we can interpret $V(r) = -1/f(r)$ as a Newtonian potential in a   spatially flat
spacetime. Hence, the genuine relativistic problem demands a non-constant $h(r)$, which
implies in a spatially curved
spacetime,  preventing the direct use of any
Newtonian result  in the present case. In particular, there is no place here for the introduction of the
so-called Henon  variables and, consequently,   the identification of the parabolic properties of the
 isochrone potentials, see \cite{PPD}.
For further details on the interpretation of the effective potential (\ref{unov}) on spatially
curved manifolds, see \cite{esp}. 
}

   The isochrony condition 
$\frac{\partial T}{\partial \ell} = 0$ is equivalent to demand $\frac{\partial \Theta}{\partial E} = 0$
since we have
$
T = 2\frac{\partial \mathcal{A}_{r}}{\partial E}$ and $
\Theta = -\frac{1}{2}\frac{\partial \mathcal{A}_{r}}{\partial \ell}$, 
with $\mathcal{A}_{r}$ standing for the so-called radial action of the problem
\begin{equation}
\mathcal{A}_{r}(E,\ell) =  2\int_{r_{\rm min}}^{r_{\rm max}} \frac{\sqrt{E-
U_{ {\ell}} (r)} dr}{\sqrt{h(r)}}.
\end{equation}
 {Notice that, as in the Newtonian case \cite{RP}, the isochrony condition for
static spherically
symmetrical spacetimes
 is fully
equivalent to demand that the relativistic radial action be additively separable, {\em i.e.},
$\mathcal{A}_{r}(E,\ell) = \mathcal{B}(E) + \mathcal{C}(\ell)$. In other words, we have
$\frac{\partial T}{\partial \ell} = 0$, and consequently  $\frac{\partial \Theta}{\partial E} = 0$, if and only if the radial action is additively separable
in the constants $E$ and $\ell$. Unfortunately, however, due to the generic presence of spatial curvature   (non constant $h(r)$),
the relativistic isochrony condition is not enough to single out relativistic isochrone potentials as it happens in the Newtonian case. }

 {
For the sake of notation simplicity, we will drop hereafter the $\ell$ index
denoting dependence in $U(r)$.  
It is convenient now to follow \cite{Perlick} and 
introduce}   a different integration variable $R$ such that
 \begin{equation}
 \label{change}
\frac{dR}{R^2} = \frac{dr}{r^2\sqrt{h}},
\end{equation}
in terms of which the azimuthal angle (\ref{apsidal}) reads
\begin{equation}
\label{apsidalR}
 \Theta(E,\ell) = {\ell}  \int_{R_{\rm min}}^{R_{\rm max}} \frac{dR}{R^2\sqrt{E-U(R)}} ,
\end{equation}
where the effective potential is now given by
\begin{equation}
\label{Ueff}
U(R) = \ell^2 v(R) - w(R),
\end{equation}
with 
 {
\begin{equation}
\label{vR}
v(R) = \frac{1}{r^2(R)},     \quad 
w(R) = \frac{1}{f(r(R))},
\end{equation} } 
\noindent and
\begin{equation}
\label{h(r)}
h(r) = \left(\frac{rv'R^2}{2}\right)^2 = \left(\frac{R^2}{r^2R'} \right)^2,
\end{equation}
 {where the prime $'$ denotes the derivative of the function for its
considered variable. 
Despite equation (\ref{apsidalR}) in the new radial variable $R$ is formally identical to the equivalent Newtonian
one, the effective potential (\ref{Ueff}) is quite different. In particular, the term corresponding
to the centrifugal barrier now is an arbitrary function and not a fixed $R^{-2}$ term as we would have in the
Newtonian case. It is clear that the problem in General Relativity  is much less restricted and that,
in principle, many other solutions are indeed possible, exactly in the same way of the
Bertrand spacetime problem.}
We stress that the spacetime functions $v(R)$ and $w(R)$ are  independent of the trajectory initial conditions and, hence, cannot depend on   $\ell$ nor $E$. Also, the condition $R'(r)>0$ tacitly assumed in (\ref{change}) and its direct consequence $v'(R)<0$ from (\ref{vR})
   will be important to select the correct solutions in the subsequent analysis. 
   In terms of the new radial variable $R$, the metric (\ref{metric}) is given by
\begin{equation}
\label{metric1}
g_{ab}dx^adx^b = - \frac{dt^2}{w(R)} + \frac{ dR^2}{\left(R^2v(R) \right)^2} + \frac{d\Omega^2}{v(R)}.
\end{equation}

\begin{figure}[t]
\begin{center}
\input{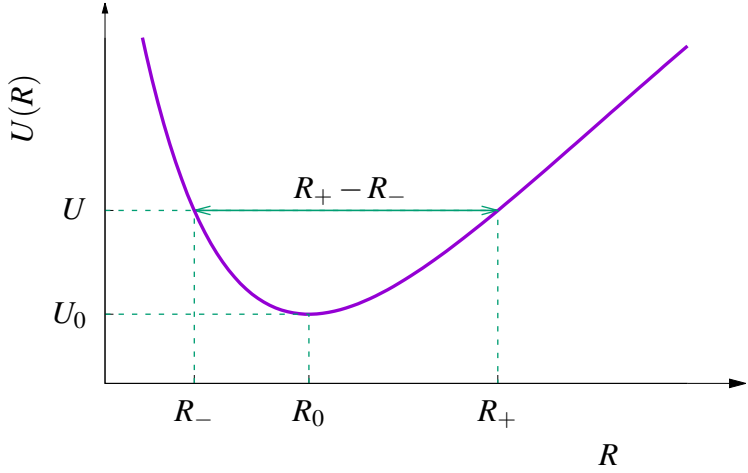}
\end{center}
 \caption{ {Aspect of a generic effective potential $U(R)$ near its local minimum at $R=R_0$. The potential
 is inverted in each of the two branches
 $R_-(U)\le R_0 $ and $R_+(U)\ge R_0$. It is clear that $R_+ - R_-$ must be a function of $U$.
 However, in the vicinity of $R_0$, $U$ is well described by a parabola  since we assume $U''_0 = U''(R_0)>0$ and,
consequently,  $ R_+ - R_- = \frac{2\sqrt{2}}{\sqrt{U''_0}}\sqrt{U - U_0} $ near $R_0$, which is the main
motivation for the proposed expression (\ref{RR1}).
 }}
 \label{fig1}
\end{figure}

Let us now decompose the motion range $[R_{\rm min},R_{\rm max}]$ into the branches $R_-\le R_0  = R(r_0)$ and $R_+\ge R_0$, where $U(R)$ in inverted in each of these branches, see Fig. \ref{fig1} and
\cite{Perlick}.  
We can write (\ref{apsidalR}) as 
\begin{equation}
\label{apsidal1}
 \Theta  =  {\ell} \int_{U_0}^{E} \frac{1}{ \sqrt{E-U }} \frac{d }{dU}
\left(\frac{1}{R_-} -   \frac{1}{R_+}
\right)dU,
\end{equation}
where $U_0=U(R_0)$. 
The isochrony condition corresponds to require 
an azimuthal angle independent of $E$, {\em i.e.}, $ \Theta(E,\ell) = \frac{\pi}{\beta_\ell}$.  Eq. (\ref{apsidal1})  can be inverted by using the Abel equation \cite{Perlick}, leading in this
case   to
\begin{equation}
\label{RR}
\frac{1}{R_-} -   \frac{1}{R_+} = \frac{ {2}}{\ell\beta_\ell}\sqrt{U-U_0}.
\end{equation}
We follow now the same approach  \cite{AJP} we have used recently for deriving the classical
isochrone potentials (\ref{he}), (\ref{bo}), and (\ref{ho}). By inspecting the analytical properties of the potential 
$U(R)$ near its minimal at $R=R_0$, see Fig. \ref{fig1}, we can write
\begin{equation}
\label{RR1}
R_+ - R_- = \frac{\sqrt{U-U_0}}{F(U)},
\end{equation}
where $F(U)$ is an arbitrary function such that $F(U_0) = \frac{\sqrt{U''_0}}{2\sqrt{2}}$,
with $U''_0=U''(R_0)$.
Notice that since $F(U)$ is assumed to be arbitrary, there is indeed no loss of generality
in choosing the form (\ref{RR1}) for the function corresponding to 
$R_+(U) - R_-(U)$. Nevertheless, this choice is a very convenient one since, from (\ref{RR}) and
(\ref{RR1}), we will have
\begin{equation}
\label{expr}
\sqrt{U-U_0} = F(U)R_+ - \frac{A}{R_+} = -\left( F(U)R_- - \frac{A}{R_-} \right),
\end{equation}
implying finally
\begin{equation}
\label{exprf}
U-U_0 =  \left( F(U)R - \frac{A}{R} \right)^2,
\end{equation}
valid for both branches, with 
\begin{equation}
A=\frac{\ell \beta_\ell}{2}.
\end{equation}
From (\ref{RR}) and
(\ref{RR1}) we have also the useful relation
\begin{equation}
\ell^2\beta^2_\ell = \frac{R_0^4U''_0}{2}. 
\end{equation}
In summary, an isochrone effective potential $U(R)$  must obey the algebraic relation (\ref{exprf})
for a function $F(U)$ such that (\ref{RR1}) holds. Hence, we indeed have a procedure to generate 
isochrone effective potentials:  once  a function $F(U)$ is given, we can formally
solve (\ref{exprf}) for $U$ and obtain the corresponding isocrone effective potential. However, of course,
we cannot always get  explicit solutions with the required properties for this equation. Fortunately, we can do it for some algebraic choices for $F(U)$. Let us consider now these explicit relevant examples.

\subsection{First case: constant $F(U)$ }
\label{secconst}
The simplest choice is a constant $F(U)=\alpha$. From (\ref{exprf}), we have in this case
simply 
\begin{equation}
\label{linear}
U(R) = BR^2 + \frac{C}{R^2} +D, 
\end{equation}
with $B=\alpha^2$, $C=A^2 = \ell^2\beta^2_\ell/4$,  and $U_0=  \alpha \ell\beta_\ell - D$. The parameter
$\alpha$ is completely arbitrary and could depend, in principle, also on $\ell$ and, hence, the decomposition of of $U(R)$ in
   ``centrifugal''  and    ``central potential'' parts as (\ref{Ueff}) can involve
the first two terms of (\ref{linear}) rather arbitrarily.  The most general
solution corresponds to set
\begin{equation}
\label{vr}
v(R) = B_1R^2 + \frac{C_1}{R^2} + D_1,
\end{equation}
and
\begin{equation}
\label{wr}
w(R) = B_2R^2 + \frac{C_2}{R^2}+ D_2.
\end{equation}
Of course, we are exploring an underlying linear structure of the 
problem, we are considering $v(R)$ and $w(R)$ as generic linear combinations of the three terms
of  (\ref{linear}) such that $B=B_1\ell^2+B_2$, $C= C_1\ell^2+C_2$, and $D=D_1\ell^2 +D_2$. 
This will be valid for all cases considered here. 
 Once $v(R)$ is given, 
we can recast the original spherical coordinates (\ref{metric}) by using 
(\ref{vR}) and   (\ref{vr}). Assuming $K=4B_1C_1\ne 0$, we will have
\begin{equation}
R^2 = \frac{1-D_1r^2 \mp \sqrt{(1-D_1r^2)^2 - Kr^4}}{2B_1r^2},
\end{equation}
leading to
\begin{equation}
h = \frac{2C_1\left((1-D_1r^2)^2 - Kr^4\right)}{1-D_1r^2 \pm \sqrt{(1-D_1r^2)^2 - Kr^4}}.
\label{bertr2hh}
\end{equation}
Notice that  (\ref{bertr2h}) and (\ref{bertr2hh}) are identical, up to the definition
of the constant $C_1$. For the determination of $f(r)$, we will explore the linear structure of the problem and consider
$w(R) =   w(R) -\xi v(R)  + \xi/r^2$, where (\ref{vR}) was invoked, with $\xi = C_2/C_1$.
Hence, without loss of generality, one can consider instead  the function
(\ref{wr}) its equivalent form
 \begin{equation}
\label{wr1}
w(R)   =   {B'_2}{R^2}+ D'_2 + \frac{\xi}{r^2},
\end{equation} 
with  $B_2' = B_2 -   \xi B_1  \ne 0 $  and  $D'_2 = D_2 - \xi D_1 $,  
 leading finally to
\begin{equation}
\label{bertr2ff}
 {\frac{1}{f}} =  G\mp \frac{\rho r^2}{1-D_1r^2 \pm \sqrt{(1-D_1r^2)^2 - Kr^4}} + \frac{\xi}{r^2}, 
\end{equation}
 where $G=D_2'$ and  $\rho = 2B'_2C_1$. The extra $r^{-2}$ term in (\ref{wr1}) and (\ref{bertr2ff}) can  be also understood
 as a manifestation of the 
  well-known gauge invariance of the problem: if $U(R)$ is an isochrone potential, then
$\tilde U(R) = U(R) + \Delta + \Lambda v(R)$, for any constants $\Delta$ and $\Lambda$, will be also isochrone, but with the
parameters   $E$ and
$\ell$  redefined accordingly.  
  The spacetime functions (\ref{bertr2hh}) and
 (\ref{bertr2ff}) define our first class of isochrone spacetimes, for which we have
 \begin{equation}
 \beta_\ell^2 = 4C_1\left( 1 + \frac{\xi}{\ell^2} \right).
 \end{equation}
 The Newtonian limit for this family is obtained by setting  $D_1 = K = 0$ and $C_1=1$, and a simple inspection of (\ref{bertr2ff}) reveals that it corresponds to the harmonic potential plus an extra $r^{-2}$ interaction term.  The Bertrand spacetimes of the second type arise by considering  the  case $\xi = 0$ 
 and a rational $\sqrt{C_1}$. The extra $r^{-2}$ term is   associated with a spacetime singularity at $r=0$ as one can see by noticing that $f \sim r^{2}$ for $r\to 0$ and $\xi \ne 0$, and
 that the scalar curvature of the metric (\ref{metric}) in this case reads
 \begin{equation}
 \mathcal{R} \sim    \frac{2-3\left(rh' + 2h \right)}{r^2} 
 \end{equation}
 for $r \to 0$, diverging for the two possibilities of (\ref{bertr2hh}). The presence of this singularity
for $\xi\ne 0$ is generic for all families of isochrone spacetimes discussed here.
 
 Let us now inspect the particular cases with $K=0$. First, of course,
one cannot have both $B_1 = 0$ and $C_1 = 0$ simultaneously, since it would correspond to no centrifugal barrier at all, which is incompatible with (\ref{vR}). 
For the case $B_1 = 0$ and $C_1\ne 0$, the expressions (\ref{bertr2hh}) and (\ref{bertr2ff})  are
still valid   in the limit $B_1\to 0$, choosing the pertinent signs.
The case $B_1\ne 0$ and $C_1=0$ is also possible but
somehow distinct, since it does not have a Newtonian limit. It corresponds to a second family of isochrone spacetimes for which 
\begin{equation}
\label{hh3}
h = \frac{(1-D_1r^2)^3}{B_1r^4}
\end{equation}
and
\begin{equation}
\label{ff3}
 {\frac{1}{f}} = G + \frac{\rho r^2}{1-D_1r^2} + \frac{\xi}{r^2}.
\end{equation}
  For this family,  we have
\begin{equation}
\beta_\ell^2 = \frac{4\rho}{B_1\ell^2}.
\end{equation}
Since $\rho\ne 0$, otherwise $U(R)$ would not have a local minimum,   there is no limit
of constant $\beta_\ell$ for these spacetimes and, consequently, they do not have a Bertrand limit. 
The clear possibility of having $h(r_*)=0$, with $r_*>0$, for $D_1>0$ in (\ref{hh3})  (and also in (\ref{bertr2hh}))  and its implication for the causal
structure of the underlying spacetime  deserve some comments. We will address these points in the
last section. 
Finally, notice the condition $B'_2\ne 0$ assumed
in (\ref{wr}) is necessary by a rather subtle reason. If $B'_2 = 0$, the effective potential has only
the centrifugal barrier term, and the requirement of a local minimum  $U(R)$ will be equivalent
of a local minimum of $v(R)$, but due to (\ref{h(r)}), we will have $h(r_0)=0$. We will also return to this point
in the last section.

\subsection{Second case: linear $F(U)$ }
\label{seclinear}

The second simplest choice for our problem is, of course, the  linear $F(U)$ case. 
 Since we can always add a constant to $U$, one can consider without loss of generality $F(U) = \alpha U$. We will proceed
along the same lines of the  preceding case 
and solve (\ref{exprf}). We get
\begin{equation}
\label{Ulin}
U(R) =  \frac{B\sqrt{p  + \epsilon R^2}}{R^2} + \frac{C}{R^2} +D,
\end{equation}
where $U_0 = -\epsilon \alpha^2B^2 - D$, with $\epsilon$ assuming two possible values $\epsilon = \pm 1$. The other parameters in the potential (\ref{Ulin}) are such that
\begin{eqnarray}
\label{Ulinp}
p&=&  \frac{4A\alpha +1}{4\alpha^4B^2},\\
\label{UlinC}
C &=& \frac{2A\alpha +1}{2\alpha^2},
\end{eqnarray}  
  and it is clear that one requires $p>0$ for $\epsilon = -1$.
The signs of $B$ and $C$ must also be conveniently chosen to guarantee a local minimum for the
effective potential $U(R)$.
One can advance from (\ref{Ulin}) that   
  the flat space limit ($R=r$)  of $U(R)$ will comprehend  all  Henon's isochrone potentials (\ref{he}),
  (\ref{bo})  and
  (\ref{ho}), but occasionally   with some extra  $r^{-2}$ terms. As in the preceding case, let us first consider 
  the linear combinations
  \begin{equation}
\label{vr2}
v(R) =    B_1 \frac{\sqrt{p + \epsilon R^2}}{R^2}  + \frac{C_1}{R^2} + D_1,
\end{equation}
and
\begin{equation}
w(R) = B'_2 \frac{\sqrt{p  + \epsilon R^2}}{R^2} + D'_2 + \frac{\xi}{r^2},
\end{equation}
with $C_1\ne 0$ and $B'_2\ne 0$.
  Reintroducing    the original spherical coordinates, we have
  \begin{equation}
R^2 =  \epsilon C_1\left( \left(\frac{  r +\mu \Phi(r)    }{ \Xi(r)  -\epsilon  \nu r} \right)^2 -\mu\right),
\end{equation}
where $\mu = p/C_1$, $\nu = B_1/2\sqrt{C_1}$,
\begin{equation}
\Phi(r) =  \epsilon\left( \frac{1}{r}  - D_1 r\right),  
\end{equation}
and
\begin{equation}
\Xi^2(r) =  \epsilon +\kappa r^2 + \mu \Phi^2(r)  ,  
\end{equation}
with $
\kappa = \nu^2- \epsilon D_1,
$
leading to
  \begin{equation}
  \label{hlin}
h = \frac{ \epsilon C_1  r \left(r+ \mu   \Phi(r) + 2\epsilon \nu \mu \left( \Xi(r) - \epsilon\nu r\right)    \right) \Xi^2(r)}{  \left( r + \mu\Phi(r) \right)^2}
\end{equation}
and
  \begin{equation}
  \label{flin}
 {\frac{1}{f}} = G + \frac{\epsilon\rho\left(  r +\mu \Phi(r)    \right)\left(  \Xi(r)  - \epsilon \nu r\right) }{\left(  r +\mu \Phi(r)    \right)^2 - \mu \left(  \Xi(r)  - \epsilon \nu r\right)^2} + \frac{\xi}{r^2},
\end{equation}
The general expression for the azimuthal angle of the metric with the functions (\ref{hlin}) and (\ref{flin}) is rather
complicated. 
For $\mu\ne 0$, we have  from (\ref{Ulinp})
and (\ref{UlinC})
\begin{equation}
\alpha^2 = \frac{1}{2C_1\zeta_\ell}\left(1-\sqrt{1-\zeta_\ell} \right),
\end{equation}
with
\begin{equation}
\zeta_\ell = \mu \left(2\nu + \frac{ \rho }{\ell^2 + \xi} \right)^2,
\end{equation}
leading   to
\begin{equation}
\label{bb3}
\beta_\ell^2 = 2C_1\left( 1 + \frac{\xi}{\ell^2} \right) \left(\frac{1}{\zeta_\ell} - 1 \right)\left(
1-\sqrt{1-\zeta_\ell}
 \right). 
\end{equation}
The case   $\mu=0$  follows straightforwardly  from the $\mu\to 0$ limit of (\ref{bb3}).  

The Newtonian limit of this family of isochrone spacetimes corresponds to
set $\kappa = \nu = 0$ (locally flat condition) and $C_1=0$ (no conic singularity at the origin), and
it corresponds to the classic Henon's potentials (\ref{he}), (\ref{bo}), and (\ref{ho}), with 
an extra $r^{-2}$ interaction. Furthermore, the parameter $\xi$ can be properly chosen to reproduce (\ref{he}), (\ref{bo}), and (\ref{ho}) exactly, eliminating the singularity at the origin. In this sense,
this family of isochrone spacetimes can be considered the relativistic version of the Henon's original
potentials. The Bertrand limit corresponds to the case $\mu = \xi = 0$ and  a rational $\sqrt{C_1}$, and
coincide with the Perlick first family defined by (\ref{bertr1f}) and (\ref{bertr1h}). The family defined by
the functions  
(\ref{hlin}) and (\ref{flin}) has some asymptotically flat spacetimes, they correspond to the cases
with $\Xi$ constant for large $r$, which demands $\epsilon=1$ and $ \mu D_1^2 + \kappa = \mu D_1^2 -   D_1 + \nu^2 = 0$, and in contrast with the Bertrand family  defined by (\ref{bertr1f}) and (\ref{bertr1h}),
these asymptotically flat isochrone spacetimes can admit curved spatial sections. 

As in the preceding case, we are left with the $C_1=0$ case, which does not have a Newtonian limit. We have 
for this case
\begin{equation}
h = \frac{B_1r\left(\sqrt{B_1^2r^2+4p \Phi^2} -\epsilon B_1r  \right)\left(B_1^2r^2+4p \Phi^2 \right)}{8p\Phi^2},
\end{equation}
\begin{equation}
 {\frac{1}{f}} = G + \frac{\rho \left(  \sqrt{B_1^2r^2+4p \Phi^2} -\epsilon B_1r\right)}{2p r} + \frac{\xi}{r^2},
\end{equation}
and
\begin{equation}
\beta_\ell^2 =   \frac{C_2}{\ell^2}  \left(\frac{1}{\bar \zeta_\ell} - 1 \right)\left(
1-\sqrt{1-\bar\zeta_\ell}
 \right),
\end{equation}
 with
$
 \bar\zeta_\ell = p(\ell^2 +\xi)^2/\rho^2.
$
This family does not admit asymptotically flat spacetimes or Bertrand limit.

\subsection{Third case: quadratic $F(U)$}

Our last explicit case is also the   next natural one: the quadratic $F(U)$. Again, thanks to the gauge invariance of the problem, we can
 consider without loss of generality $F(U) = \alpha( U^2 + \gamma)$. Equation (\ref{exprf}) in this case
 will be an irreducible fourth-order polynomial, but it admits the simple solution
 \begin{equation}
 U(R) = B\sqrt{\frac{1}{R} + p} + \frac{C}{R} + D,
 \end{equation}
 where $\alpha^2 = \frac{1}{4B^2C}$, $\gamma = -B^2p$, $U_0 = -\frac{B^2}{4C} -Cp -D$, and
 \begin{equation}
 A = \frac{C^\frac{3}{2}}{2B}.
 \end{equation}
 The corresponding spacetimes are not asymptotically flat and have no Newtonian limit. We have for
 $C_1\ne 0$:
 \begin{eqnarray}
 h  &=&  \frac{C_1^2r^2  \Omega^2  }{4\left(B_1 + \Omega \right)^2} , \\
 {\frac{1}{f}}& =& G + \rho \sqrt{ \Omega^2 + B_1\left(B_1 + 2\Omega \right) } + \frac{\xi}{r^2},
 \end{eqnarray}
  with
 \begin{equation}
 \Omega^2 = B_1^2 + 4C_1\left( C_1 p  -D  +  \frac{1}{r^2}\right),
 \end{equation}
 and
 \begin{equation}
 \beta_\ell^2 = \frac{\sqrt{C_1}\left( \ell^2 + \xi \right)^{\frac{3}{2}}}{ \ell^2\left(2\rho+  \frac{B_1}{C_1}\left(\ell^2 + \xi\right) \right) }.
 \end{equation}
For case with $C_1=0$, we have
\begin{eqnarray}
 h  &=&  \left(\frac{B_1^2r^3   }{4\left(1-D_1r^2 \right)} \right)^2, \\
 {\frac{1}{f}}& =& G + \frac{\rho\left( \left(1-D_1r^2 \right)^2 -B_1^2p \right)}{B_1^2r^4} + \frac{\xi}{r^2},
 \end{eqnarray}
 and
 \begin{equation}
 \beta_\ell^2 = \frac{\rho^\frac{3}{2}}{2B_1^2(\ell^2+\xi)}.
 \end{equation}
These families do not admit asymptotically flat spacetimes and have no
Bertrand limit.

\section{Final Remarks}

\begin{figure}
\tikzstyle{naveqs} = [ draw , text centered, text width=3.5cm,  rounded corners]

\resizebox{.48\textwidth}{!}{
\begin{tikzpicture}
\matrix[row sep=4cm,column sep=1cm, ampersand replacement=\&] {%
    \node (p1) [naveqs] {Isochrone spacetime $F(U)=\alpha$};   \& \&  \node (p2) [naveqs] {Bertrand spacetime \\ of the second type };  \\
        \& \node [naveqs](p3) {Harmonic potential}; \&    \\
};
\draw   (p1) edge [->,>=stealth,shorten <=2pt, shorten >=2pt,thick] node[above]{Closed orbits}  (p2) ;
\draw 	(p1) edge [->,>=stealth,shorten <=2pt, shorten >=2pt,thick] node[sloped, above]{Newtonian limit}  (p3) ;
\draw   (p2) edge [->,>=stealth,shorten <=2pt, shorten >=2pt, thick] node[sloped, above]{Newtonian limit} (p3);
     \end{tikzpicture}
}
\resizebox{.48\textwidth}{!}{
\begin{tikzpicture}
\matrix[row sep=4cm,column sep=2.8cm ,  ampersand replacement=\&] {%
    \node (p1) [naveqs] {Isochrone spacetime $F(U)=\alpha U$};   \& \&  \node (p2) [naveqs] {Bertrand spacetime \\ of the first type };  \\
          \node [naveqs](p3) {Hen\'on potentials}; \&  \& \node [naveqs](p4) {Newtonian potential};  \\
};
\draw   (p1) edge [->,>=stealth,shorten <=2pt, shorten >=2pt,thick] node[above]{Closed orbits}  (p2) ;
\draw 	(p1) edge [->,>=stealth,shorten <=2pt, shorten >=2pt,thick] node[sloped, above]{Newtonian limit}  (p3) ;
\draw   (p2) edge [->,>=stealth,shorten <=2pt, shorten >=2pt, thick] node[sloped, above]{Newtonian limit} (p4);
\draw   (p3) edge [->,>=stealth,shorten <=2pt, shorten >=2pt,thick] node[above]{Closed orbits}  (p4) ;
\end{tikzpicture}}
\caption{\label{fig2} {
Schematic representation of the physically most relevant isochrone spacetimes. Left: the constant $F(U)$ 
family of section \ref{secconst}. It has the  Bertrand   spacetimes of the second type
(\ref{bertr2f}) and (\ref{bertr2h}) as its limit with closed orbits, and both have the harmonic
potential as their Newtonian limit. Right: The linear  $F(U)$ 
family of section
\ref{seclinear}. Its particular cases with  closed orbits are   the
Bertrand   spacetimes of the first type
(\ref{bertr1f}) and (\ref{bertr1h}), and its respective Newtonian limit corresponds to the Hen\'on isochrone potentials (\ref{he}),
(\ref{bo}) and (\ref{ho}). The Newtonian potential arises as the Newtonian limit of the
Bertrand   spacetimes of the first type and as the closed orbits case of the isochrone potentials.
}}
\end{figure}
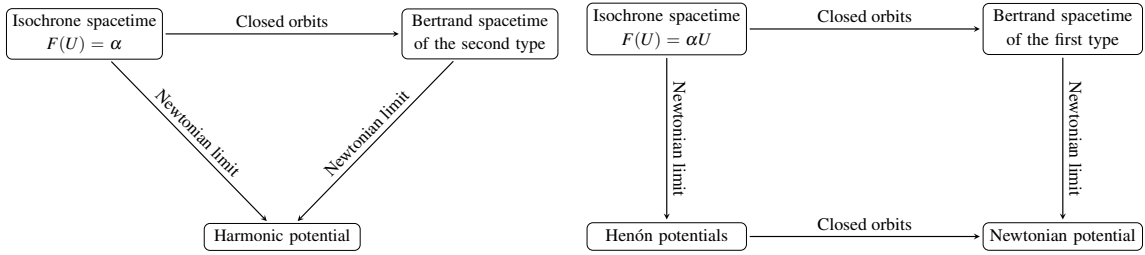

We have introduced the notion of   isochrone spacetimes and presented a procedure to generate
them by means of the algebraic equation (\ref{exprf}). We have obtained explicit
expressions for the families of spacetimes corresponding to the cases with constant, linear, and quadratic $F(U)$, but many others solutions are indeed possible.
We remind that we are looking for solutions of (\ref{exprf}) allowing for, at least, a three-dimensional 
vector space   that will
give origin to 
  the two  functions $v(R)$ and $w(R)$ of the effective potential (\ref{Ueff}). With the help of Maple,
we could find explicit   solutions for (\ref{exprf})  with the required properties for $F(U) = \alpha\sqrt{U}+\gamma$, $F(U) = \alpha U^{-1}+ \gamma$, and $F(U) = \frac{\alpha U}{U+\gamma}$, among others, but the resulting expressions are too cumbersome to be useful in our context.  For cases  with
 $F(U)$ involving 
any rational expressions with non-linear polynomials,    
the problem of solving (\ref{exprf}) reduces to find  the roots of higher order irreducible polynomials in $U$, and in some cases the solutions are indeed compatible with our requirements, despite 
their typical intricate expressions. We do not expect any compatible solution for non-algebraic functions $F(U)$, but we could not prove that they indeed do not exist.
 {Nevertheless, the constant and linear $F(U)$ cases are particularly relevant here because they have the Bertrand
spacetimes \cite{Perlick} as special limits, see Fig. \ref{fig2}.}
 {
 Moreover, the linear case has as  
 Newtonian limit the   Henon’s isochrone potentials (\ref{he}),
(\ref{bo}) and (\ref{ho}), and, hence, such family of isochrone spacetimes could be useful as  a relativistic extension of the original Henon's proposal for the study of globular clusters. They
might also generalize some recent models for galactic dark matter  \cite{astroB} based in the Bertrand spacetimes of the
first type. It is worth stressing that only the constant and linear $F(U)$ cases have isochrone 
Newtonian limits and, hence, they are probably the most relevant isochrone spacetimes.}

The causal structure of isochrone spacetimes certainly deserves a deeper investigation, but some
issues are already clear. One can see from all explicit families of spacetime of the previous section that
one can have,  for some choice of parameters, $h(r_*)=0$ with $r_*>0$. In such a case, from the metric (\ref{metric}), we see that  $r=r_*$  is a null sphere and, hence, a candidate to be an event horizon provided no spacetime
singularity is present at
$r=r_*$. Incidentally, this is the reason why one cannot have $B_2'=0$ in the discussions of the last section, since in
such a case we would have $r_0 = r_*$, a circular null orbit, and our analysis is focused from the beginning  only on
timelike geodesics. 
From (\ref{h(r)}), we see that $r_*$ corresponds to a critical
point of $v(R)$, and from (\ref{metric1}) we see that no evident spacetime singularity is expected at the
critical points of $v(R)$, provided $w(R)$ be regular there. Hence, despite that $R$ defined in (\ref{change}) is a tortoise-like
coordinate,  there should be a possibility  of extending the isochrone spacetime across $r_*$, giving
origin to a black hole spacetime in the case of absence of singularities  at  $r =r_*$. The possibility of
having an isochrone black hole is certainly instigating from a theoretical point of view. As an illustration, let us consider the case with $C_1=0$ for a constant $F(U)$, which corresponds to the
functions (\ref{hh3}) and (\ref{ff3}), with the parameters   $G=\xi=0$ and $\rho^\frac{1}{2} = B_1^\frac{1}{4} = D_1^\frac{1}{2} = r_*^{-1}$   for the sake of simplicity,
\begin{equation}
\label{new}
g_{ab}dx^adx^b = -\frac{r_*^2-r^2}{r^2}dt^2 + \frac{r_*^2r^4 }{(r_*^2- r^2)^3} dr^2 + r^2d\Omega^2.
\end{equation} 
Both the scalar curvature and the Kretschmann invariant suggest that the null suface $r=r_*$ is
not singular, and hence the metric (\ref{new}) could be extended naturally for $r>r_*$. Of course, this metric
is not static in this exterior region, and hence our discussion on isochrone orbits does not apply there.  
On the other hand, we can also see form the curvature invariants  that there is a naked spacetime singularity in the interior region at $r=0$.

We finishing noticing that the isochrone spacetimes generically violates the weak energy condition, as one can see
from the Einstein tensor evaluated for the metric (\ref{metric}). We have for the temporal component
\begin{equation}
G_{00} = \frac{f(r)\left( 1 -\frac{d}{dr} rh(r)   \right)}{r^2},
\end{equation}
and it is clear that one can have $G_{00}<0$ for sufficiently large $r$ for all families we have
considered in this paper. This raises a pertinent question about the physical interpretation of the isochrone spacetimes, since in the context of General Relativity they would require exotic matter. Nevertheless, in the context
of modified gravity, the energy conditions can be considerably different from those ones of General Relativity \cite{EC}, and perhaps some of these
theories could have isochrone spacetimes as physically viable solutions. 

\section*{Acknowledgment}
AS acknowledges the financial support of
CNPq and FAPESP (Brazil) through the grants  302674/2018-7  and  21/09293-7, respectively, 
 and thanks   Vitor Cardoso and Jos\'e S. Lemos  for 
 the warm hospitality at the Center for Astrophysics and Gravitation of the
 University of Lisbon, where this work was finished.
 
 \section*{Data availability}
This paper has no associated data.

\end{document}